\documentclass[conference]{IEEEtran}
\IEEEoverridecommandlockouts

\usepackage{cite}
\usepackage{amsmath,amssymb,amsfonts}
\usepackage[ruled,vlined,linesnumbered]{algorithm2e}
\usepackage{graphicx}
\usepackage{subcaption} 
\usepackage{textcomp}
\usepackage{enumitem}
\usepackage{pgfplots}
\usepackage{pgfplotstable}
\pgfplotsset{compat=1.18} 
\usepackage{tikz}
\usepackage{array} 
\usepackage[percent]{overpic}

\usepackage{amssymb}
\usepackage{pifont}
\usepackage{hyperref}
\newcommand{\cmark}{\ding{51}}%
\newcommand{\xmark}{\ding{55}}%

\SetKwInput{KwData}{Input}                 
\SetKwInput{KwResult}{Output}              
\SetKwComment{Comment}{/* }{ */}           
\SetAlgoNlRelativeSize{-1}                 
\SetNlSty{textbf}{(}{)}                    
\SetAlgoNlRelativeSize{-1}                 
\DontPrintSemicolon                        
\SetAlgoVlined  
\usepackage{xcolor}  
\definecolor{customgreen}{rgb}{0.835, 0.91, 0.831}
\definecolor{customred}{rgb}{0.98, 0.8, 0.8}
\definecolor{customblue}{rgb}{0.85, 0.91, 0.98}
\definecolor{walletcolor}{RGB}{31, 119, 180}   
\definecolor{votingcolor}{RGB}{44, 160, 44}    
\definecolor{mixedcolor}{RGB}{255, 127, 14}  
\SetCommentSty{textnormal}
\SetSideCommentLeft
\SetNlSty{textbf}{}{}  

\newcommand{\cmnt}[1]{}
\newcommand{\ignore}[1]{}
\newcommand{\remove}[1]{}

\newcommand{\floor}[1]{\lfloor #1 \rfloor}

\newcommand{\secref}[1]{Section~\ref{sec:#1}}
\newcommand{\figref}[1]{Fig.~\ref{fig:#1}}

\newcommand{\algoref}[1]{{Algorithm \ref{alg:#1}}}

\usepackage{xcolor}
\def\BibTeX{{\rm B\kern-.05em{\sc i\kern-.025em b}\kern-.08em
    T\kern-.1667em\lower.7ex\hbox{E}\kern-.125emX}}
    
 \makeatletter

\newcommand{\linebreakand}{%
  \end{@IEEEauthorhalign}
  \hfill\mbox{}\par
  \mbox{}\hfill\begin{@IEEEauthorhalign}
}
\makeatother
\begin{document}

\title{BlockRaFT: A Distributed Framework for Fault-Tolerant and Scalable Blockchain Nodes\thanks{This work is partly funded by Meity Project No.4(4)/2021-ITEA and a research grant from SUI foundation}}

\author{\IEEEauthorblockN{Manaswini Piduguralla}
\IEEEauthorblockA{\textit{Indian Institute of Technology Hyderabad,} India \\
cs20resch11007@iith.ac.in}
\and
\IEEEauthorblockN{Souvik Sarkar}
\IEEEauthorblockA{\textit{Indian Institute of Technology Hyderabad,} India \\
cs23mtech02001@iith.ac.in}
\linebreakand 
\hspace*{10mm}
\IEEEauthorblockN{\hspace{-1mm}Arunmoezhi Ramachandran}
\IEEEauthorblockA{\hspace{10mm}\textit{\ Independent Researcher,} India \\
\hspace*{7mm}arunmoezhi@gmail.com}
\and
\hspace*{5mm}
\IEEEauthorblockN{Sathya Peri}
\IEEEauthorblockA{\hspace{10mm}\textit{Indian Institute of Technology Hyderabad,} India \\
\ \ \ \ \ \ \ \ \ sathya\_p@cse.iith.ac.in}
}

\maketitle

\begin{abstract}

Blockchain technology enhances transparency by maintaining a distributed ledger among mutually untrusting parties. Despite its advantages, scalability and availability remain critical bottlenecks that hinder widespread adoption. The increasing complexity of blockchain nodes further necessitates robust fault tolerance and high throughput to ensure seamless operations.

We present BlockRaFT\footnote{Code is available at: https://github.com/PDCRL/SCT-DistFramework.git}, a crash-tolerant distributed framework designed to improve both the scalability and reliability of blockchain node operations. BlockRaFT framework utilizes RAFT consensus protocol to elect a leader within a cluster of systems. The elected leader coordinates and distributes workloads across follower nodes, thereby optimizing resource utilization and work load balancing. We analyzed the tasks performed by blockchain nodes and partition them according to their stateful and stateless characteristics. Stateless operations are centralized at the leader, while stateful operations are replicated and coordinated across the cluster to ensure consistency and fault tolerance. We evaluate whether this distributed intra-node architecture provides measurable benefits over traditional single-node execution models in terms of scalability, availability, and performance.

Additionally, we introduce a concurrent Merkle tree optimization that decouples smart contract execution from tree updates, significantly reducing one of the significant performance overheads in blockchain systems.
Our design philosophy is rooted in utilizing the well-established principles of distributed computing and customizing them for the blockchain domain rather than reinventing them.

\end{abstract}


\section{Introduction}
\label{sec:intro}
Satoshi Nakamoto introduced blockchain technology as the foundation for the cryptocurrency Bitcoin \cite{Nakamoto:Bitcoin:2009}. A decentralized ledger system that allows multiple parties to maintain a shared, immutable record of transactions without depending on a central authority. The system is built on well-established principles of cryptography. These cryptographic foundations ensure data integrity, enable secure transactions, and build trust in a network of untrusted participants.
Initially developed for digital currencies, blockchain has evolved into a general-purpose distributed ledger technology. Its applications now extend far beyond cryptocurrency. Blockchain is being explored and adopted in various fields, including supply chain management, healthcare, real estate, identity management, and finance \cite{Bodkhe+:IEEEAccess:Review:2020}. These use cases leverage blockchain’s core features decentralization, transparency, immutability, and auditability.

A permissioned blockchain is a network with restricted access to approved participants \cite{Androulaki+:Hyperledger:Eurosys:2018}. One or more organizations govern these networks. They control who can join the network and what actions each participant can perform. In contrast, public blockchains such as Bitcoin or Ethereum are open to anyone and are maintained by decentralized communities.
Permissioned blockchains offer several advantages for regulatory settings. Permissioned blockchains combine the blockchain framework's security, trust, and immutability properties with the necessary security and environmental control features that organizations need.

\vspace{1mm}
\noindent
\textbf{Drawbacks of Current Frameworks.}
Despite these benefits, scalability and availability remain the key challenges in permissioned blockchains. Blockchain networks often suffer from lower throughput and higher latency than traditional centralized systems \cite{Dickerson+:ACSC:PODC:2017,parwat:springer:2021, Manaswini+:2023:europar}. Even in well-optimized systems, performance may lag by several orders of magnitude. Another limitation arises from the node architecture in permissioned settings. Organizations are typically limited in the number of nodes they operate. This is because the cost of running blockchain protocols increases with network size. Protocols such as gossip communication and consensus mechanisms become significantly more expensive as more nodes are added. Consensus protocols like Practical Byzantine Fault Tolerance (PBFT) \cite{gilad+:SOSP:Algorand:2017} and Proof of Stake (PoS) \cite{Vasin:2014:WP} increase exponentially with an increase in the number of nodes.

Although the blockchain network is resilient to node failures, individual organizations may lose access if their nodes go offline. In such cases, the network continues to function. However, the affected organization cannot submit transactions, query the blockchain, or validate blocks until its nodes are restored, as illustrated in \figref{dist}. Node crashes pose significant challenges in a permissioned blockchain. As blockchain adoption expands, researchers and developers are actively working on improving scalability, fault tolerance, and interoperability.




\noindent
\textbf{Contributions.} This paper makes the following key contributions:
\begin{enumerate}[label=(\alph*)]
    \item We design a crash-tolerant and scalable distributed framework  for permissioned blockchain nodes based on a RAFT leader-follower cluster model that eliminates single points of organizational failure in \secref{model}.
    \item A three-phase concurrent Merkle tree optimization that decouples smart contract execution from state tree updates is introduced \secref{model}.
    \item We provide the algorithmic design of the proposed framework (\secref{implementation}) and conduct extensive evaluations to validate its performance and effectiveness (\secref{experiments}).
\end{enumerate}

\section{Background}
\label{sec:background}
\subsection{Blockchain}
Blockchain \cite{Nakamoto:Bitcoin:2009} is a distributed ledger system where changes to the ledger are made through transactions packed in blocks. Each block contains a set of transactions and is cryptographically linked to its predecessor through a hash of the previous block's contents. This chaining of blocks through hash ensures the integrity of the data, as altering a block's contents would require modifying all subsequent blocks. Blocks often contain multiple smart contract transactions (SCTs), which are self-executing pieces of code that execute the terms of an agreement between two or more parties. A network of nodes maintains this distributed ledger. Each node maintains a full or partial copy of the distributed ledger and participates in verifying and propagating new transactions and blocks. Agreement on a block being added to the chain is accomplished through various consensus mechanisms, like proof of work (PoW)~\cite{Nakamoto:Bitcoin:2009}, proof of stake (PoS)~\cite{Vasin:2014:WP}, and proof of elapsed time (PoET)~\cite{KunBla+:1997:CJ}.

\subsection{Merkle Tree}
A Merkle tree \cite{Liu+2021:EIECS:markleTree} is a cryptographic data structure widely used in blockchain systems to verify transactions and state information efficiently. In a Merkle tree, each leaf node holds the cryptographic hash of a piece of data (such as a transaction), while every non-leaf (internal) node contains the hash of the concatenated hashes of its child nodes. This hierarchical hashing structure means that even a single-bit change in any leaf node will propagate up the tree, altering the Merkle root at the top. As a result, the Merkle root serves as a compact and reliable fingerprint of all the data beneath it. By comparing just the Merkle roots, blockchain nodes can quickly and securely verify that they share the same data without exchanging the entire dataset.

\subsection{RAFT Protocol}
RAFT \cite{raft} is a crash-tolerant consensus algorithm developed as a more understandable alternative to Paxos \cite{lamport:2001:paxos}. It is designed to manage a replicated log across multiple nodes in a distributed system, ensuring consistency and fault tolerance. RAFT implements consensus through leader election, and the elected leader is responsible for the log management. Raft divides time into terms of arbitrary length, and each term begins with an election. A candidate wins an election if they receive votes from a majority of the voters in the same term. The votes are cast based on the first-come, first-served (FIFO) order. The majority rule ensures that at most one candidate can win the election for a particular term. We utilize the RAFT protocol for leader election in our proposed framework.

\section{Proposed Framework}
\label{sec:model}
The proposed model introduces a distributed framework for blockchain nodes to improve two critical aspects of blockchain networks: scalability and fault tolerance. Blockchain networks suffer from lower throughput and higher latency compared to traditional centralized systems. Although blockchain networks are inherently designed to tolerate crashes and Byzantine faults at the network level, the reliability of individual nodes remains a concern. These issues must be addressed before blockchain networks become feasible alternatives for large-scale applications.

Denial of service due to node crashes are particularly pronounced in permissioned blockchain networks, where each node represents an organization. In such scenarios, the failure of an organization's node can result in the organization's operations being halted (\figref{dist}). A simple solution would be for an organization to maintain multiple nodes representing itself in the network. This approach has two significant drawbacks. Increasing the number of nodes in the network, increases the message complexity. And increase in message complexity directly impacts the overall performance of the blockchain network. Secondly, this approach compromises the fairness of blockchains. If one organization maintains more nodes than others, it gains disproportionate representation in the consensus process. Therefore, adding more full nodes per organization is not a viable or fair solution.
\subsection{BlockRaft Design}
\begin{figure}[h]
  \centering
        \centering
        \includegraphics[scale=0.4]{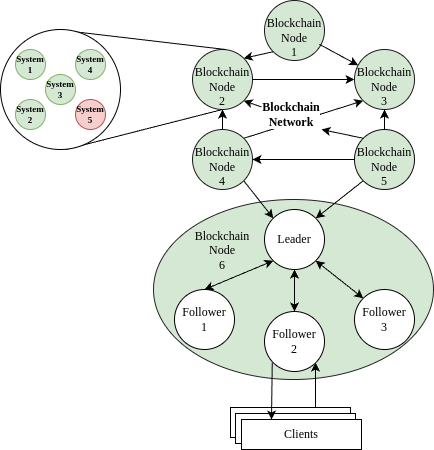}
        \caption{Distributed Framework Design for the Blockchain Node.}
	\label{fig:dist}
\end{figure}

BlockRaft framework is designed such that it would not impact the network message complexity or skew the fairness. As illustrated in \figref{dist}, a cluster of systems represents the blockchain node (node 2) in the network, and even if one or more nodes crash, the blockchain node still performs the operations. Our proposed BlockRaFT architecture is a leader-follower approach, and the node acting as a leader will represent the network. All the nodes are identical, and a leader is elected using the RAFT consensus. If a follower node crashes, the leader will redistribute the crashed follower's work to another node, and if the leader crashes, another node is elected as the new leader, and it starts performing the leader's duties. As we use RAFT consensus for leader election, the tolerance to crashes is $\floor{(n-1)/2}$ when n nodes exist in the cluster.
In our approach to the blockchain network, each node still looks like a single, unified entity. This helps preserve the principle that every organization has equal representation. Inside the node, though, the workload is shared across multiple systems, which allows it to scale out without affecting its role in the network. This framework improves performance and ensures that client access is not restricted if part of the node fails.

\begin{figure*}
    \centering
    \begin{overpic}[scale=0.6]{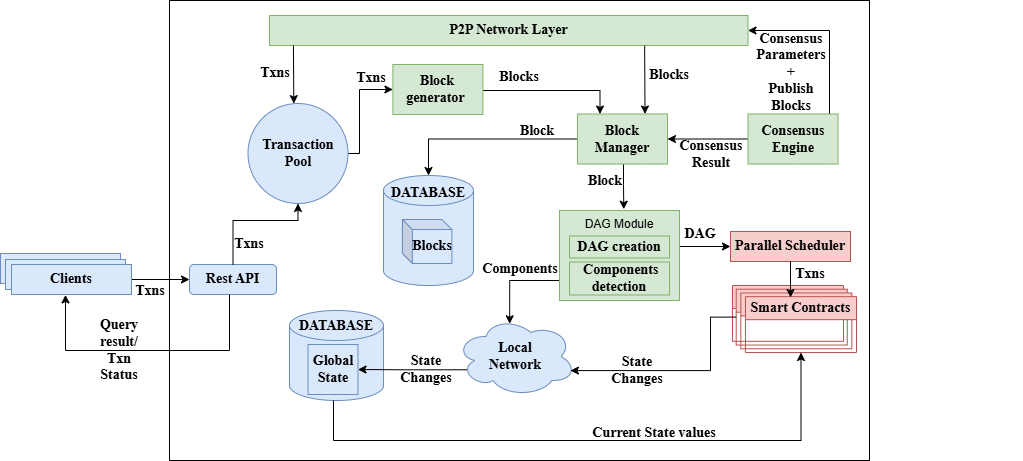}
        \put(85,10){
            \begin{tikzpicture}[scale=0.5]
                \fill[customgreen] (0,0) rectangle (0.5,0.5);
                \node[anchor=west, text=black] at (0.4,0.15) {Only Leader};
            \end{tikzpicture}
        }

        \put(85,6){
            \begin{tikzpicture}[scale=0.5]
                \fill[customred] (0,0) rectangle (0.5,0.5);
                \node[anchor=west, text=black] at (0.4,0.15) {Only Follower};
            \end{tikzpicture}
        }

        \put(85,2){
            \begin{tikzpicture}[scale=0.5]
                \fill[customblue] (0,0) rectangle (0.5,0.5);
                \node[anchor=west, text=black] at (0.4,0.28) {Both};
            \end{tikzpicture}
        }
    \end{overpic}
    \caption{Node Architecture in the Proposed Framework}
    \label{fig:NodeArch}
\end{figure*}
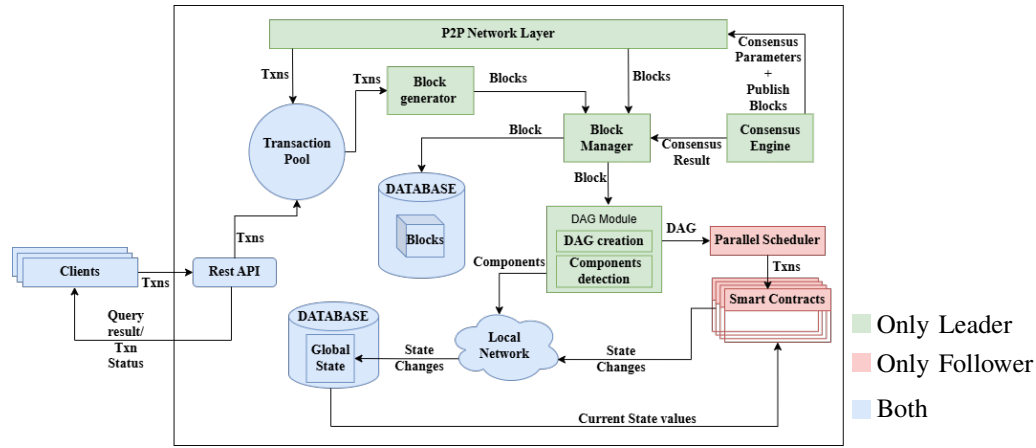

Here, in \figref{NodeArch} we present a detailed breakdown of the node architecture, and outline each module's specific roles and responsibilities in both leader and follower nodes. The architecture comprises several modules, like the network layer, Block Producer, DAG Module, Consensus Engine, and REST API. Each module is responsible for a distinct set of tasks, from managing internal state to interacting with clients. We group these modules into stateful and stateless categories. The leader handles all stateless operations, while stateful operations are maintained by every node, including the leader. The associated workload is shared across the cluster by the leader. The legend in \figref{NodeArch} clarifies which components remain active under different roles, providing a comprehensive overview of system operations.

\noindent
\textbf{Node Module: }
The Node module is responsible for initializing and managing the role-based behavior of each node within the distributed blockchain framework. Each node participates in the RAFT consensus algorithm to elect a leader upon startup. Depending on its designated role, leader or follower, the node executes a corresponding set of tasks. Follower nodes periodically monitor the leader's status to see if the leader changes or crashes and wait for task assignments. Leader nodes initiate block production or validation based on the current blockchain state. Additionally, each node maintains a health monitoring thread to ensure reliability and responsiveness within the cluster.

\noindent
\textbf{Leader Module}
The Leader module coordinates the core blockchain operations. The leader node interacts with the network and represents the blockchain node to the network. During block production, the leader gathers pending transactions from the transaction pool, verifies them, and constructs a directed acyclic graph (DAG) of the dependencies among the transactions. It decomposes the DAG into independent components, assigns them to followers, and tracks their execution. The leader aggregates the resulting state changes, finalizes the block, and publishes it to the blockchain. In validation mode, the leader ensures that the computed state from worker nodes matches the expected final state embedded in the block.

\noindent
\textbf{Follower Module}
The Follower module executes transactional workloads delegated by the leader. Each follower reconstructs the DAG based on information received from the leader and executes its assigned components in parallel according to the DAG's topological ordering. Followers report completed execution states and synchronize changes across the cluster. They also respond to state commit or discard instructions following block validation.

\noindent
\textbf{Block Producer}
The Block Producer aggregates transactions from the shared transaction pool and constructs new blocks. The transaction pool is a shared distributed queue maintained among all nodes in the cluster. It performs essential preprocessing steps like transaction validation, signature verification, and expiration filtering. The resulting block is then forwarded to the Block Manager for further processing.

\noindent
\textbf{Block Manager}
The Block Manager acts as the central coordinator for block lifecycle management. It receives newly created blocks from the Block Producer and blocks from other nodes for validation. These blocks are then dispatched to the DAG Module for conflict analysis and decomposition. The Block Manager also interfaces with the Consensus Engine to confirm the consensus outcome and ensures that only validated blocks progress to final state computation and storage.

\noindent
\textbf{DAG Module}

To facilitate efficient transaction execution, our framework incorporates a Directed Acyclic Graph (DAG) module, which operates in two distinct phases:

\paragraph{DAG Construction}  
The framework models inter-transaction dependencies using an adjacency matrix that captures read-write (RW), write-read (WR), and write-write (WW) relationships. Upon receiving a block of transactions, the DAG module employs multithreaded processing to analyze each transaction's read and write addresses. Dependencies are represented as directed edges in the adjacency matrix, where edges from transactions with lower IDs to those with higher IDs. This technique enforces acyclicity, a crucial property for maintaining deterministic execution order. The DAG module is inspired by the parallel execution model proposed in \cite{Manaswini+:2023:europar}. We have adapted the module for our distributed approach.

\paragraph{Component Detection}  
A Disjoint Set Union (DSU)~\cite{Tarjan+:1975:ACM} structure is used to detect connected components in the graph efficiently. Each node is initially placed in its own set, and the find function with path compression ensures that every node can quickly locate its root representative. When an edge is found between two nodes in the adjacency matrix, the unite operation merges their sets using union.

\noindent
\textbf{Parallel Scheduler}
The Parallel Scheduler is responsible for scheduling transaction execution based on the DAG structure provided by the leader. The scheduler identifies the transactions belonging to its assigned components for each follower and schedules them for parallel execution. An indegree array is maintained to manage dependencies, recording the number of incoming edges (dependencies) for each transaction. The scheduler continuously searches for transactions with zero indegree, i.e., those with no unmet dependencies, and dispatches them for execution. After the transaction execution, the indegree values of all dependent transactions (i.e., those with outgoing edges from the completed transaction) are decremented. The scheduler ensures that transactions are scheduled in parallel, enabling efficient parallelism.

\noindent
\textbf{REST API}
The REST API is the gateway for client interaction with the blockchain system. It handles incoming requests for read-only queries and state-changing transactions (SCTs). Read-only requests are directly resolved from the global state, while SCTs are stored in the shared transaction pool for inclusion in future blocks.

\noindent
\textbf{Network Layer}
The Network Layer facilitates communication between the blockchain framework and external blockchain networks. It is responsible for relaying transactions to the transaction pool, forwarding blocks, and transmitting consensus information. The network layer is active in the node acting as the leader in the cluster.

\noindent
\textbf{Consensus Engine}
The Consensus Engine executes the chosen consensus protocol (e.g., Proof of Work, PBFT) to validate and approve new blocks. It ensures agreement among participating nodes on the correctness and ordering of blocks. The leader node participates in consensus on behalf of the cluster.

\subsection{Concurrent Merkle Tree Optimization}
The global state is stored in most blockchain systems using a Merkle tree. This data structure is used to ensure that any state changes can be verified consistently across all nodes in the network. The root of the Merkle tree, known as the Merkle root, is included in each block and serves as a cryptographic summary of the entire state. However, they naturally introduce update dependencies: modifying any leaf node requires updating all of its ancestors to the root. As a result, any two updates to the tree are inherently dependent, making it challenging to execute smart contracts in parallel while maintaining a consistent state. To address this bottleneck, we propose a three-phase optimization strategy:
\subsubsection{Concurrent Execution with Hash Map}
During transaction execution, instead of writing changes directly to the global state (i.e., the Merkle tree), we read the required state data but store any modifications in a concurrent hash map. This structure allows multiple threads to read from the global state while independently recording changes. Since no updates are made to the Merkle tree at this stage, parallel execution becomes feasible. 

\subsubsection{Parallel Application to Leaf Nodes}
Once all transactions have been executed, each node in the cluster exchanges its recorded state changes with the others. The block's complete set of state updates is then applied in parallel, but only to the leaf nodes of the Merkle tree. Because each leaf node is modified exactly once (based on the final, post-transaction state), there is no contention, making parallel application efficient and safe.
\subsubsection{Sequential Tree Recalculation}
The modified leaf nodes' internal nodes (parents and ancestors) are updated in the final phase. This step is performed sequentially to compute the new Merkle root. This sequential recalculation is significantly less expensive than updating during transaction execution.

By decoupling transaction execution from Merkle tree updates and structuring the process into these three phases, we significantly reduce the overhead associated with Merkle tree operations. 

\subsubsection{Impact of Crashes}

This subsection discusses the safety and crash-consistency implications of the proposed three-phase Merkle tree optimization.

\paragraph{Phase 1: Execution and Local State Recording.}
During transaction execution, all transactions read state values from the committed global Merkle tree. Instead of directly updating the tree, modifications are recorded in a concurrent hash map local to each follower. The DAG module guarantees that no two transactions assigned to different followers operate on overlapping addresses within the same block. Consequently, transactions observe a consistent snapshot of the global state, and read-after-write hazards within a block are avoided. Throughout Phase 1, the global Merkle tree remains unchanged and fully consistent.

\paragraph{Phase 2: Parallel Leaf Updates.}
After all transactions in the block complete execution, each follower shares its recorded state changes with the leader. The leader aggregates these updates and distributes the combined state changes across the cluster. The leaf nodes corresponding to the modified keys are then updated in parallel. Because each key is updated exactly once based on the final execution result, no contention or inconsistency arises during this phase.

\paragraph{Phase 3: Parent Hash Recalculation.}
Once all leaf updates are applied, the internal nodes of the Merkle tree are recomputed sequentially along the affected paths to produce the new Merkle root. The block is considered committed only after this recomputation completes successfully.

\paragraph{Leader Failures.}
If the leader crashes at any point before Phase 3 completes, the block is not committed and the global state remains unchanged. Upon election of a new leader via RAFT, the cluster re-synchronizes and re-executes the block deterministically from the last committed state. Since intermediate updates are not externally visible, no inconsistent state is exposed.

\paragraph{Follower Failures.}
If a follower crashes during Phase 1, the leader detects the failure via heartbeat monitoring and reassigns the corresponding transaction components to another active follower. If a follower crashes after sharing its local state changes, the aggregated updates are already preserved by the leader, and execution proceeds with the remaining quorum.

\paragraph{Block-Level Atomicity.}
As in traditional sequential Merkle tree implementations, the block is treated as the atomic unit of state transition. The Merkle tree is not externally accessed during block execution; only the committed root after Phase 3 becomes visible. Therefore, the proposed optimization does not introduce new consistency windows compared to a sequential design. Instead, it reduces the duration of state recalculation while preserving atomicity and correctness guarantees.
\section{Algorithm Design}
\label{sec:implementation}
In the BlockRaft distributed framework, the Node Protocol (\algoref{Node-Protocol}) serves as the general lifecycle controller for each node in the cluster, determining whether a node acts as a leader or a follower at any given time. At initialization, each node participates in a consensus mechanism ( Raft protocol) to elect a leader. Once elected, the leader node activates the Leader Protocol to coordinate transaction execution. All other nodes default to the Follower Protocol and await instructions and updates from the current leader. 

\begin{algorithm}[]
\small
\caption{Node Protocol}\label{alg:Node-Protocol}
\tcp{Initialize leader and follower objects}
\texttt{leader} $\leftarrow$ \texttt{leaderObj}, \texttt{follower} $\leftarrow$ \texttt{followerObj}\;

\tcp{Start threads to monitor cluster health}
\texttt{Monitor} $\leftarrow$ \texttt{startThread(clusterMonitorFunc)}\;
\tcp{Run Raft protocol until successful}
\While{\texttt{not raftSuccess()}}{
    \texttt{runRaftProtocol()}\;
}
\tcp{Execute node protocol as long as the cluster is healthy}
\While{\texttt{clusterHealth.load()}}{
    \If{\texttt{IsLeader()}}{
        \tcp{Start the recurring lease on the Leader key}
        \texttt{leaderLease} $\leftarrow$ \texttt{startThread(leaderHeartbeat)}\;

        \tcp{Run leader protocol to produce or validate blocks}
        \texttt{leader.leaderProtocol()}\;
    }
    \Else{
        \tcp{Start a thread to monitor the status of leader}
        \texttt{leaderMonitor} $\leftarrow$ \texttt{startThread(leaderMonitorFunc)}\;
        \tcp{Run follower protocol}
        \texttt{follower.followerProtocol()}\;
    }
}

\tcp{Wait for monitor thread to complete}
\texttt{clusterMonitorFunc.join()}\;

\If{\texttt{IsLeader()}}{
    \tcp{Wait for leader lease thread if node is a leader}
    \texttt{leaderHeartbeat.join()}\;
}

\Return\;
\end{algorithm}

The Leader Protocol (\algoref{Leader-Protocol}) orchestrates the execution by collecting a new block of transactions from the network, generating a DAG from those transactions, and partitioning the DAG into independent components. These components represent a set of transactions that can be executed in parallel without conflicts. The leader assigns these components to available follower nodes based on the current cluster status, ensuring a quorum of active followers is available before proceeding. After broadcasting the component assignments and a start signal, the leader monitors the progress and completion of execution, eventually collecting state updates from the followers.

\begin{algorithm}[]
\small
\caption{Leader Protocol}\label{alg:Leader-Protocol}
\tcp{Initialize DAG module}
\texttt{DAGmodule} $\leftarrow$ \texttt{DAGObj}\;

\tcp{Get the latest block for validation from the network layer}
\texttt{latestBlock} $\leftarrow$ \texttt{NetworkLayer.getBlock()}\;

\tcp{Get the total member list of the cluster}
$n \leftarrow$ \texttt{getClusterMembers()}\;

\tcp{Get the list of active followers currently}
$k \leftarrow$ \texttt{getActiveFollowers()}\;

\If{$k < \left\lfloor \frac{n}{2} \right\rfloor + 1$}{
    \Return\;
}

\tcp{Create DAG from the transactions in the block}
\texttt{DAG} $\leftarrow$ \texttt{DAGmodule.create(latestBlock)}\;

\tcp{Derive the independent component list from the DAG}
\texttt{componentList} $\leftarrow$ \texttt{DAGmodule.connectedComponents(DAG)}\;

\tcp{Assign followers for each component in the list}
\texttt{assignFollowers(componentList)}\;

\tcp{Start a thread to monitor followers' status}
\texttt{followersMonitor} $\leftarrow$ \texttt{startThread(monitorFollowers)}\;

\tcp{Share the block and components list with followers}
\texttt{sendFollowers(latestBlock)}\;
\texttt{sendFollowers(componentList)}\;

\tcp{Inform the followers to start execution}
\texttt{sendFollowers("start")}\;

\tcp{Check periodically the component execution status}
\texttt{checkComponents(componentList)}\;

\tcp{Inform the followers that execution is complete}
\texttt{sendFollowers("finish")}\;

\tcp{Collect state changes from the followers}
\texttt{saveData(memberList)}\;

\tcp{Wait for threads to join and clean the objects}
\texttt{followersMonitor.join()}\;
\texttt{DAGmodule.dagClean()}\;
\texttt{componentList.clean()}\;
\Return\;
\end{algorithm}

On the other hand, the Follower Protocol (\algoref{Follower-Protocol}) runs on each follower node and acts upon the data received from the leader. Each follower retrieves the current block and component assignments on startup and waits until the leader instructs them to begin execution. Once the “start” signal is received, followers execute their assigned components using a local scheduler object, which manages the parallel transaction execution. Meanwhile, they also monitor for new component assignments, allowing dynamic reassignment or load balancing. After completing their work, followers notify the leader and wait for a “finish” instruction, indicating either global completion or more work to be assigned.

\begin{algorithm}[]
\small
\caption{Follower Protocol}\label{alg:Follower-Protocol}
\tcp{Initialize scheduler object}
\texttt{schedulerObj} $\leftarrow$ \texttt{scheduler()}\;

\tcp{Start a thread to watch for leader crashes}
\texttt{leaderMonitor} $\leftarrow$ \texttt{startThread(monitorLeader)}\;

\tcp{Get the block and components list from the leader}
\texttt{latestBlock} $\leftarrow$ \texttt{receive(block)}\;
\texttt{componentList} $\leftarrow$ \texttt{receive(components)}\;

\tcp{Wait for leader's instruction to start execution}
\While{\texttt{receive(runKey)} $\ne$ ``start"}{
    \texttt{/* wait */}
}

\tcp{Start monitoring for new component assignments}
\texttt{receive(watchComponentChanges)}\;

\tcp{Execute transactions of assigned components}
\texttt{schedulerObj.execute(componentList)}\;

\tcp{Inform leader about execution completion}
\texttt{sendLeader(componentID.status, ``done")}\;

\tcp{Wait for leader's final instruction}
\While{\texttt{receive(runKey)} $\ne$ ``finish"}{
    \texttt{/* check for new work or wait */}
}

\tcp{Collect state changes from other followers}
\texttt{sendData(ClusterList)}\;

\tcp{Join monitor thread and cleanup}
\texttt{leaderMonitor.join()}\;
\texttt{schedulerObj.clean()}\;

\Return\;
\end{algorithm}


\subsection{Crash Handling}
Fault tolerance in the BlockRaFT framework is built on the RAFT leader election algorithm. When a node starts, it immediately participates in RAFT to elect a leader. Once the leader is chosen, all nodes employ a heartbeat mechanism to signal their liveness. A leader’s heartbeat confirms its authority for the current term. Within the leader module, a dedicated thread continuously monitors follower heartbeats, while in each follower, a thread tracks the leader’s activity.

If the leader detects a follower crash and the number of failed nodes remains less than $(n-1)/2$, the work assigned to the crashed follower is reassigned to another follower, ensuring continuity. However, if crashes exceed $(n-1)/2$, the system halts all operations to preserve consistency. In case a follower detects the leader’s failure, the cluster immediately triggers a new RAFT leader election to restore coordination.

\subsection{Concurrent Merkle Tree}

Our proposed algorithm aims to enable high-throughput transaction execution in blockchain systems while preserving the consistency guarantees provided by Merkle trees. \algoref{concurrentMerkleTree} presents the design of our concurrent Merkle tree operations. Instead of updating the tree directly during execution, the algorithm separates the process into distinct phases. This separation eliminates redundant work and allows safe parallelism. We have \texttt{updateValue} function, which records modifications in a concurrent hash map rather than directly updating the Merkle tree. By acquiring an accessor and writing the new value into \texttt{myMap}, multiple threads can apply updates concurrently without introducing conflicts. This ensures that transaction execution is decoupled from immediate tree modification, enabling parallelism. The retrieval operations \texttt{getRootHash}, \texttt{getValue}, and \texttt{getNode} provide efficient ways to query the state. The \texttt{getRootHash} function extracts the current Merkle root from persistent storage, while \texttt{getValue} and \texttt{getNode} retrieve values or deserialized nodes from the database. These lightweight functions do not interfere with concurrent updates, ensuring that read operations can occur alongside transaction execution.

The \texttt{parallelInsertFromMap} procedure implements the second phase of the optimization strategy. After execution, all recorded key–value pairs in the concurrent hash map are distributed across multiple threads in balanced chunks. Each thread inserts its assigned updates into the leaf nodes of the Merkle tree in parallel. Since every leaf node is updated exactly once based on the final state, this step avoids contention and maximizes throughput. Finally, once all leaf updates are complete, the algorithm performs a sequential recomputation of the Merkle tree’s internal nodes using the \texttt{updateParentHashes} procedure. The ancestors up to the root are updated for each modified key to reflect the new leaf values. This ensures the correctness of the Merkle root while keeping the recomputation cost low because only affected paths are traversed. The \texttt{updateParentHashes} and \texttt{getRootHash} functions are implemented in the same manner as in a serial Merkle tree; therefore, we omit their details from the algorithm due to space constraints.

\begin{algorithm}[h]
\small
\caption{ConcurrentMerkleTree Operations}
\label{alg:concurrentMerkleTree}
\SetKwFunction{FUpdateValue}{updateValue}
\SetKwFunction{FGetRootHash}{getRootHash}
\SetKwFunction{FParallelInsertFromMap}{parallelInsertFromMap}
\SetKwFunction{FGetValue}{getValue}
\SetKwFunction{FGetNode}{getNode}

\textbf{Data Structure:} \texttt{myMap} $\gets$ concurrent hash map \;

\SetKwProg{Fn}{Function}{:}{}
\SetKwProg{Pn}{Procedure}{:}{}

\Fn{\FUpdateValue{key, value}}{
    Acquire accessor \texttt{acc} on \texttt{myMap} \;
    Insert $(key)$ into \texttt{myMap} with \texttt{acc} \;
    $\texttt{acc.second} \gets value$ \;
}

\Fn{\FGetValue{key}}{
    $keyHash \gets$ \texttt{computeHash}(key) \;
    $node \gets$ \FGetNode{keyHash} \;
    \Return $node.value$ \;
}

\Fn{\FGetNode{key}}{
    $data \gets$ read from database using $key$ \;
    \uIf{$data$ exists}{
        \Return \texttt{data} \;
    }
        \Return empty Node \;
}
\Fn{\FParallelInsertFromMap{$nThreads$}}{
    $items \gets$ all key-value iterators from \texttt{myMap} \;
    $totalItems \gets$ size of $items$ \;
    $chunkSize \gets \lceil totalItems / nThreads \rceil$ \;
    \For{$i \gets 0$ \KwTo $nThreads-1$}{
        $start \gets i \times chunkSize$ \;
        $end \gets \min(start + chunkSize, totalItems)$ \;
        Launch thread \;
        \For{$j \gets start$ \KwTo $end-1$}{
            $(key, value) \gets items[j]$ \;
            \texttt{insert}(key, value) \;
        }
    }
    Join all threads \;
    \ForEach{$(key) \in$ \texttt{myMap}}{
        \texttt{updateParentHashes}(key) \;
    }
}
\end{algorithm}

\section{Experimental Results}
\label{sec:experiments}
\subsection*{Experimental Setup}

\textbf{Operating System:} Ubuntu (64-bit)
\textbf{Processor:}
\begin{itemize}
    \item Model: AMD EPYC 7452 32-Core Processor
    \item Architecture: x86\_64
    \item CPU Count: 128 logical CPUs 
        \item Memory: 251 GB, Buffer/Cache: 7.2 GB
    \item Base Frequency: 1.5 GHz, Max Boost: 2.35 GHz
\end{itemize}
\textbf{Individual node configuration}
Operating System: Ubuntu (64-bit)
\textbf{Processor:}
\begin{itemize}
    \item CPU Count: 16 logical CPUs 
    \item Memory: 9.7 GB,  Buffer/Cache: 3.4 GB
\end{itemize}
We implemented the proposed BlockRaFT framework in C++, leveraging ETCD \cite{etcd} as a distributed asynchronous shared memory system for communication and coordination among cluster nodes. ETCD facilitates consistent state exchange and simplifies coordination, while its built-in RAFT protocol supports leader election. In our implementation, we piggybacked on the ETCD leader for efficient cluster management. In addition to the ETCD-based shared-memory communication model, we also implemented a message-passing–based communication model, whose results are presented in \secref{experiments} and \figref{conflict}. We observed that both communication methodologies perform very similarly, with negligible differences in convergence behavior. We chose ETCD for the final system due to its practical performance and ease of integration.

\subsection*{Concurrent Merkle Tree Experiments}

\noindent
\paragraph{Baseline Scope and Design Rationale.}
In our evaluation, the comparison is conducted against a traditional sequential Merkle tree implementation integrated within the same execution framework. This baseline performs updates incrementally during transaction execution, recomputing affected hashes immediately after each modification, which reflects the conventional update pattern used in many blockchain implementations.

We note that several advanced Merkle tree designs, such as Jellyfish~\cite{gao2021jellyfish} and Angela~\cite{kalidhindi2018angela} exsists but these focus on improving persistence efficiency and versioning. These approaches address important challenges in blockchain state management and are highly effective. Our approach differs as we try to take advantage of batch based updates in blockchain and optimizing through creating a separation between transaction execution and state storage.

\begin{figure*}[!t]
\centering

\begin{subfigure}{0.32\textwidth}
\centering
\begin{tikzpicture}[scale=0.6]
 \begin{axis}[
 title={Execution Time vs percentage of read operations},
        xlabel=percentage of read operations,
        ylabel style={align=center}, 
        ylabel=Parallel (sec),
        ymin=0, ymax=0.4,
        symbolic x coords={0,20,40,60,80,100},
        ymajorgrids = true,
                 grid style = dashed,
        legend style={
                        at={(0.25,-0.2)},
            anchor=north,
            legend columns=1,
               fill=none, draw=none,
        },        every axis plot/.append style = { line width = 2pt },
        /pgfplots/legend image code/.code = {%
            \draw[mark repeat=2,mark phase=2,#1] 
            plot coordinates {
                (0cm,0cm) 
                (0.7cm,0cm)
                (1.4cm,0cm)
                (2.1cm,0cm)
                (2.8cm,0cm)%
            };
        },
        nodes near coords,
    ]
\addplot[color=walletcolor, mark=o, mark options={fill=white,solid}, mark size=4pt, smooth,
                 every node near coord/.append style = {xshift = 0pt, yshift = 7pt}] coordinates {
(0,0.3387)  [0.3387]
(20,0.3467)  [0.3467]
(40,0.365)  [0.365]
(60,0.3668)  [0.3668]
(80,0.373)  [0.373]
(100,0.0729)  [0.0729]
    };
    \addlegendentry{Parallel}
 \end{axis}

 \begin{axis}[
        ylabel=Serial (sec),
        ymin=0, ymax=130,
        axis y line*=right,
        axis x line=none,
        symbolic x coords={0,20,40,60,80,100},
        enlargelimits=0.05,
        every node near coord/.append style={
            font=\small, anchor=north, yshift=0pt
        },
        legend style={
                        at={(0.9,-0.2)},
            anchor=north,
            legend columns=1,
               fill=none, draw=none,
        },        every axis plot/.append style = { line width = 2pt },
        /pgfplots/legend image code/.code = {%
            \draw[mark repeat=2,mark phase=2,#1] 
            plot coordinates {
                (0cm,0cm) 
                (0.7cm,0cm)
                (1.4cm,0cm)
                (2.1cm,0cm)
                (2.8cm,0cm)%
            };
        },
        nodes near coords,
    ]
\addplot[color=red, mark=+, mark options={fill=white,solid}, mark size=4pt, solid, 
                 every node near coord/.append style = {xshift = 0pt, yshift = -5pt}] coordinates {
(0,116.7795)  [116.7795]
(20,93.2841)  [93.2841]
(40,69.7882)  [69.7882]
(60,47.0715)  [47.0715]
(80,23.9102)  [23.9102]
(100,0.396)  [0.396]

    };
    \addlegendentry{Serial}
 \end{axis}
\end{tikzpicture}
\caption{Read/Write Percentage}
\label{fig:merkleTree-readWrite}
\end{subfigure}
\hfill
\begin{subfigure}{0.32\textwidth}
\centering
\begin{tikzpicture}[scale=0.6]
 \begin{axis}[
 title={Execution Time vs Operations Count},
        xlabel=Operations Count,
        ylabel style={align=center}, 
        ylabel=Parallel (sec),
        ymin=0.2, ymax=1.0,
symbolic x coords={20k,40k,60k,80k,100k},
           ymajorgrids = true,
                 grid style = dashed,
        legend style={
            at={(0.25,-0.2)},
            anchor=north,
            legend columns=1,
             fill=none, draw=none,
        },
        every axis plot/.append style = { line width = 2pt },
        /pgfplots/legend image code/.code = {%
            \draw[mark repeat=2,mark phase=2,#1] 
            plot coordinates {
                (0cm,0cm) 
                (0.7cm,0cm)
                (1.4cm,0cm)
                (2.1cm,0cm)
                (2.8cm,0cm)%
            };
        },
        nodes near coords,
    ]
        \addplot[color=walletcolor, mark=o, mark options={fill=white,solid}, mark size=4pt, smooth,
                 every node near coord/.append style = {xshift = 0pt, yshift = 7pt}]    coordinates {
                 
        (20k,0.3269) [0.3269]
        (40k,0.3311) [0.3311]
        (60k,0.3438) [0.3438]
        (80k,0.3468) [0.3468]
        (100k,0.3564) [0.3564]

    };
    \addlegendentry{Parallel}
 \end{axis}

 \begin{axis}[
        ylabel=Serial (sec),
        ymin=0, ymax=85,
        axis y line*=right,
        axis x line=none,
symbolic x coords={20k,40k,60k,80k,100k},
        every node near coord/.append style={
            font=\small, anchor=north, yshift=0pt
        },
        legend style={
            at={(0.9,-0.2)},
            anchor=north,
            legend columns=1,
             fill=none, draw=none,
        },
      every axis plot/.append style = { line width = 2pt },
        /pgfplots/legend image code/.code = {%
            \draw[mark repeat=2,mark phase=2,#1] 
            plot coordinates {
                (0cm,0cm) 
                (0.7cm,0cm)
                (1.4cm,0cm)
                (2.1cm,0cm)
                (2.8cm,0cm)%
            };
        },
        nodes near coords,
    ]
        \addplot[color=red, mark=+, mark options={fill=white,solid}, mark size=4pt, solid, smooth,
                 every node near coord/.append style = {xshift = 0pt, yshift = -5pt}] coordinates {
                         (20k,16.7211)  [16.7211]
        (40k,33.4602)  [33.4602]
        (60k,50.3049)  [50.3049]
        (80k,67.3562)  [67.3562]
        (100k,83.8106) [83.8106]
    };
    \addlegendentry{Serial}
 \end{axis}
\end{tikzpicture}
\caption{Operations Count}
\label{fig:merkleTree-operations}
\end{subfigure}
\hfill
\begin{subfigure}{0.32\textwidth}
\centering
\begin{tikzpicture}[scale=0.6]
 \begin{axis}[
        xlabel=Thread Counts,
        ylabel style={align=center}, 
        ylabel=Parallel (sec),
        ymin=0.2, ymax=1.0,
        xtick=data,
           ymajorgrids = true,
                 grid style = dashed,
                     enlarge x limits = 0.05,
        legend style={
            at={(0.25,-0.2)},
            anchor=north,
            legend columns=1,
             fill=none, draw=none,
        },
        every axis plot/.append style = { line width = 2pt },
        /pgfplots/legend image code/.code = {%
            \draw[mark repeat=2,mark phase=2,#1] 
            plot coordinates {
                (0cm,0cm) 
                (0.7cm,0cm)
                (1.4cm,0cm)
                (2.1cm,0cm)
                (2.8cm,0cm)%
            };
        },
        nodes near coords,
    ]
        \addplot[color=walletcolor, mark=o, mark options={fill=white,solid}, mark size=4pt, smooth,
                 every node near coord/.append style = {xshift = 0pt, yshift = 7pt}]    coordinates {
(2,0.7991)  [0.7991]
(4,0.548)  [0.548]
(8,0.4142)  [0.4142]
(16,0.3422)  [0.3422]
(32,0.2973)  [0.2973]
(64,0.4256)  [0.4256]
    };
    \addlegendentry{Parallel}
 \end{axis}

 \begin{axis}[
 title={Execution Time vs Thread Counts},
        ylabel=Serial (sec),
        ymin=0, ymax=85,
        axis y line*=right,
        axis x line=none,
       xtick=data,
        enlargelimits=0.05,
        every node near coord/.append style={
            font=\small, anchor=north, yshift=0pt
        },
        legend style={
            at={(0.9,-0.2)},
            anchor=north,
            legend columns=1,
             fill=none, draw=none,
        },
      every axis plot/.append style = { line width = 2pt },
        /pgfplots/legend image code/.code = {%
            \draw[mark repeat=2,mark phase=2,#1] 
            plot coordinates {
                (0cm,0cm) 
                (0.7cm,0cm)
                (1.4cm,0cm)
                (2.1cm,0cm)
                (2.8cm,0cm)%
            };
        },
        nodes near coords,
    ]
        \addplot[color=red, mark=+, mark options={fill=white,solid}, mark size=4pt, solid, smooth,
                 every node near coord/.append style = {xshift = 0pt, yshift = -5pt}] coordinates {
(2,82.4681)  [82.4681]
(4,82.4681)  [82.4681]
(8,82.4681)  [82.4681]
(16,82.4681)  [82.4681]
(32,82.4681)  [82.4681]
(64,82.4681)  [82.4681]
    };
    \addlegendentry{Serial}
 \end{axis}
\end{tikzpicture}
\caption{Thread Count}
\label{fig:merkleTree-threads}
\end{subfigure}

\caption{Performance comparison of Serial and Parallel execution under different workloads.}
\label{fig:combined-performanceTree}
\end{figure*}
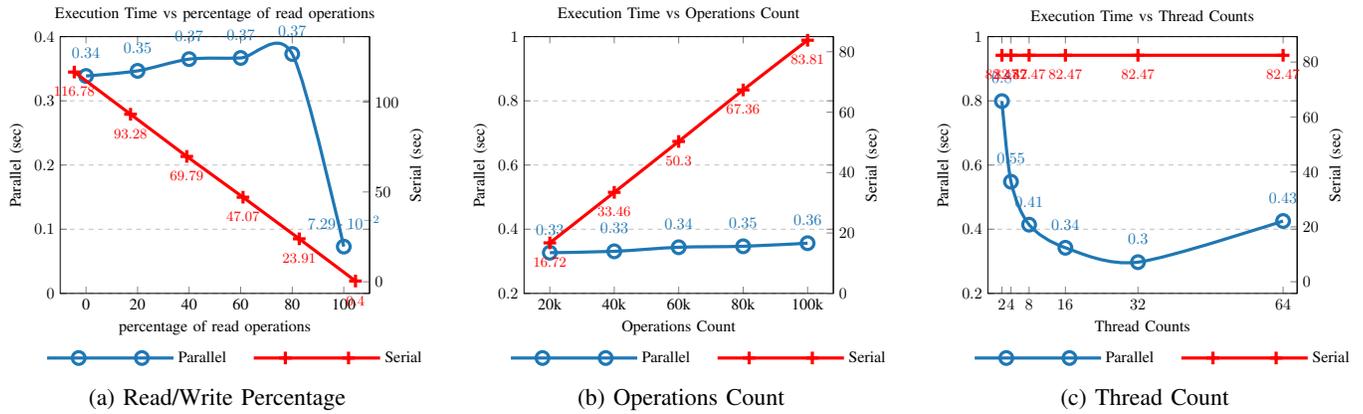

\textbf{RQ 1:}  \emph{Does the proposed concurrent Merkle tree strategy significantly reduce state update overhead compared to a sequential implementation?}

Our objective is to measure how the optimization behaves under varying workload compositions, transaction volumes, and levels of parallelism. We conduct three sets of experiments: (i) varying the read–write ratio from $0\%$ to $100\%$ in increments of $20\%$ \figref{merkleTree-readWrite}, (ii) varying the total number of operations from 20K to 100K in increments of 20K \figref{merkleTree-operations}, and (iii) varying the number of threads from 2 to 64 in powers of two \figref{merkleTree-threads}

We employed the YCSB benchmark \cite{Djellel+:OLTP-Bench:2013} suite using BenchBase repository \cite{benchbase}. Following prior work \cite{dual-plot}, we employ a dual-y-axis plot sharing a common x-axis, enabling an efficient and space-saving comparison between the serial and parallel results. 

\figref{merkleTree-readWrite} evaluates how workload composition influences performance. As expected, write-intensive workloads benefit most from the proposed design, since Merkle tree updates dominate execution cost in these scenarios. At 100\% writes, the concurrent implementation significantly outperforms the sequential baseline. Even in read-only workloads, where state updates are minimal, the design maintains approximately a 5× improvement due to parallel execution. 

\figref{merkleTree-operations} measures scalability as the number of operations increases. The concurrent Merkle tree exhibits near-linear scaling, while the sequential baseline grows proportionally with workload size. At 100K operations, the optimized version consistently outperforms the serial implementation by more than two orders of magnitude.

\figref{merkleTree-threads} evaluates parallel scalability. Execution time decreases substantially as the thread count increases up to 32 threads, indicating effective parallelization of leaf updates and state recording. At 64 threads, a slight slowdown is observed due to synchronization overhead and resource contention.

Across all experiments, the concurrent Merkle tree consistently outperforms the sequential implementation, particularly under write-heavy and large-scale workloads. The results validate that separating transaction execution from Merkle tree updates substantially reduces state update overhead while preserving correctness.

\subsection{BlockRaft Experiments}
\noindent
\paragraph{Comaprision models.}
To evaluate the performance of our proposed BlockRaft framework, we implemented two distinct comparison models within our system architecture. All configurations share identical smart contract logic, DAG construction, storage mechanisms, and Merkle tree structures to ensure that performance differences arise solely from architectural design decisions.

\begin{enumerate}
\item \textbf{Single-core Execution Model:}
This model represents the traditional blockchain processing approach. Transactions are executed sequentially on a single processing core. This model serves as a baseline for understanding the performance of conventional blockchain systems without any parallelism or workload distribution.
\item \textbf{Multi-core Execution Model:}  
This configuration executes transactions in parallel across multiple processing cores in a single system. However, unlike BlockRaft, this model does not incorporate sophisticated workload distribution or coordination mechanisms. The parallel execution benefits from concurrency and Merkle tree optimization but lacks a fault-tolerant design. This baseline isolates the benefits of intra-node distribution and fault tolerance by comparing BlockRaFT against a purely shared-memory parallel execution model.
\end{enumerate}
These baselines are intentionally implemented within the same framework to ensure controlled and fair comparison. Our goal is quantify the benefits of distributed intra-node clustering and evaluate the overhead of crash-tolerant coordination.

\textbf{RQ 2: Scalability}  \emph{Does BlockRaFT maintain near-linear execution time scaling under varying conditions, and does it outperform the baseline models?}

\textbf{RQ 3: Fault Tolerance}  \emph{Is the performance overhead of BlockRaFT's distributed coordination is a reasonable trade-off for the fault-tolerance guarantees provided?}

We evaluate the performance of BlockRaFT using two smart contracts: \textit{Voting} and \textit{Wallet}. Our objective is to analyze how execution time is affected by key parameters, including the number of transactions per block, degree of dependency, thread count, execution delay, cluster size, and fault scenarios.

The degree of dependency is defined as the percentage of actual directed edges in the transaction dependency graph (DAG) relative to the total number of possible edges. This metric quantifies the level of interdependence among transactions within a block. A higher percentage indicates a denser dependency graph, which restricts parallelism and increases coordination overhead.

For each experiment, we compare BlockRaFT against multi-core and single-core baselines to assess scalability and distributed overhead. The Voting contract models a decentralized voting system that supports voter and candidate registration, vote casting, vote transfers, and secure result queries. Additional results for the Wallet contract are available in our repository~\cite{BlockRAFT}.

\begin{figure*}[!ht]
\centering

\begin{subfigure}{0.32\textwidth}
\centering
\begin{tikzpicture}[scale=0.6]
    \begin{axis}[
    title={Threads: 16, Cluster Size: 3, Txns: 4000},
    xlabel={Conflict Percentage},
    ylabel={Execution Time (sec)},
      scaled ticks=false,
         xmin=0,
xmax=5,
    xtick=data,
     ymin=0,
ymax=12,
    ytick={0,2, 4,6,8,10,12},  
        ymajorgrids = true,
        grid style = dashed,
        enlarge x limits = 0.05,
        legend style = {
            at = {(0.5,-0.2)},
            anchor = north,
            legend columns = 1,
            fill=none, draw=none,
        },
        every axis plot/.append style = { line width = 2pt },
        /pgfplots/legend image code/.code = {%
            \draw[mark repeat=2,mark phase=2,#1] 
            plot coordinates {
                (0cm,0cm) 
                (0.7cm,0cm)
                (1.4cm,0cm)
                (2.1cm,0cm)
                (2.8cm,0cm)%
            };
        },
        nodes near coords,
    ]

        \addplot[color=walletcolor, mark=o, mark options={fill=white,solid}, mark size=4pt, 
                 every node near coord/.append style = {xshift = 0pt, yshift = -17pt}] 
            table[x=Conflict, y=BlockRAFT, col sep=space] {results/conflict.txt};
        \addplot[color=violet, mark=x, mark options={fill=white,solid}, mark size=4pt, 
                 every node near coord/.append style = {xshift = 0pt, yshift = -1pt}] 
            table[x=Conflict, y=BlockRAFT-Msg, col sep=space] {results/conflict.txt};

        \addplot[color=mixedcolor, mark=triangle*, mark options={fill=white,solid}, mark size=4pt, dashed, 
                 every node near coord/.append style = {xshift = 0pt, yshift = -18pt}] 
            table[x=Conflict, y=MultiCore, col sep=space] {results/conflict.txt};

        \addplot[color=red, mark=+, mark options={fill=white,solid}, mark size=4pt, solid,
                 every node near coord/.append style = {xshift = 0pt, yshift = 5pt}] 
            table[x=Conflict, y=SingleCore, col sep=space] {results/conflict.txt};

        \legend{BlockRaFT Voting SCT, BlockRaFT-Msg Voting SCT,  Multi-core Voting SCT,  single-core Voting SCT}

    \end{axis}	
\end{tikzpicture}
\caption{Conflict Percentage}
\label{fig:conflict}
\end{subfigure}
\hfill
\begin{subfigure}{0.32\textwidth}
\centering
\begin{tikzpicture}[scale=0.6]
    \begin{axis}[
    title={Threads: 16, Cluster Size: 3, conflict: 3\%},
    xlabel={Transactions per Block},
    ylabel={Execution Time (sec)},
      scaled ticks=false,
         xmin=1000,
xmax=5000,
    xtick=data,
     ymin=0,
ymax=14,
    ytick={0,2, 4,6,8,10,12,14},  
        ymajorgrids = true,
        grid style = dashed,
        enlarge x limits = 0.05,
        legend style = {
            at = {(0.5,-0.2)},
            anchor = north,
            legend columns = 1,
            fill=none, draw=none,
        },
        every axis plot/.append style = { line width = 2pt },
        /pgfplots/legend image code/.code = {%
            \draw[mark repeat=2,mark phase=2,#1] 
            plot coordinates {
                (0cm,0cm) 
                (0.7cm,0cm)
                (1.4cm,0cm)
                (2.1cm,0cm)
                (2.8cm,0cm)%
            };
        },
        nodes near coords,
    ]

        \addplot[color=walletcolor, mark=o, mark options={fill=white,solid}, mark size=4pt, 
                 every node near coord/.append style = {xshift = 0pt, yshift = 7pt}] 
            table[x=txns, y=BlockRAFT, col sep=space] {results/txns.txt};

        \addplot[color=mixedcolor, mark=triangle*, mark options={fill=white,solid}, mark size=4pt, dashed, 
                 every node near coord/.append style = {xshift = 0pt, yshift = -18pt}] 
            table[x=txns, y=MultiCore, col sep=space] {results/txns.txt};

        \addplot[color=red, mark=+, mark options={fill=white,solid}, mark size=4pt, solid,
                 every node near coord/.append style = {xshift = 0pt, yshift = 5pt}] 
            table[x=txns, y=SingleCore, col sep=space] {results/txns.txt};

        \legend{BlockRaFT Voting SCT,  Multi-core Voting SCT,  single-core Voting SCT}

    \end{axis}	
    \end{tikzpicture}
\caption{Transactions Count}
\label{fig:txns}
\end{subfigure}
\hfill
\begin{subfigure}{0.32\textwidth}
\centering
\begin{tikzpicture}[scale=0.6]
    \begin{axis}[
    title={Cluster Size: 3, Txns: 4000, Conflict: 5\%},
    xlabel={Thread Counts},
    ylabel={Execution Time (sec)},
      scaled ticks=false,
         xmin=0,
xmax=64,
    xtick={2,4,8,16,32,64},
     ymin=0,
ymax=12,
    ytick={0,2, 4,6,8,10,12},  
        ymajorgrids = true,
        grid style = dashed,
        enlarge x limits = 0.05,
        legend style = {
            at = {(0.5,-0.2)},
            anchor = north,
            legend columns = 1,
            fill=none, draw=none,
        },
        every axis plot/.append style = { line width = 2pt },
        /pgfplots/legend image code/.code = {%
            \draw[mark repeat=2,mark phase=2,#1] 
            plot coordinates {
                (0cm,0cm) 
                (0.7cm,0cm)
                (1.4cm,0cm)
                (2.1cm,0cm)
                (2.8cm,0cm)%
            };
        },
        nodes near coords,
    ]
        \addplot[color=walletcolor, mark=o, mark options={fill=white,solid}, mark size=4pt, smooth,
                 every node near coord/.append style = {xshift = 0pt, yshift = 7pt}] 
            table[x=Threads, y=BlockRAFT, col sep=space] {results/threads.txt};

        \addplot[color=mixedcolor, mark=triangle*, mark options={fill=white,solid}, mark size=4pt, dashed, smooth,
                 every node near coord/.append style = {xshift = 0pt, yshift = -18pt}] 
            table[x=Threads, y=MultiCore, col sep=space] {results/threads.txt};

        \addplot[color=red, mark=+, mark options={fill=white,solid}, mark size=4pt, solid, smooth,
                 every node near coord/.append style = {xshift = 0pt, yshift = 5pt}] 
            table[x=Threads, y=SingleCore, col sep=space] {results/threads.txt};
            
        \legend{BlockRaFT Voting SCT,  Multi-core Voting SCT,  single-core Voting SCT}

    \end{axis}	
\end{tikzpicture}
\caption{Thread Count}
\label{fig:threads}
\end{subfigure}

\caption{Performance comparison of Serial, Parallel and distributed blockchain frameworks under different workloads.}
\label{fig:combined-performance}
\end{figure*}
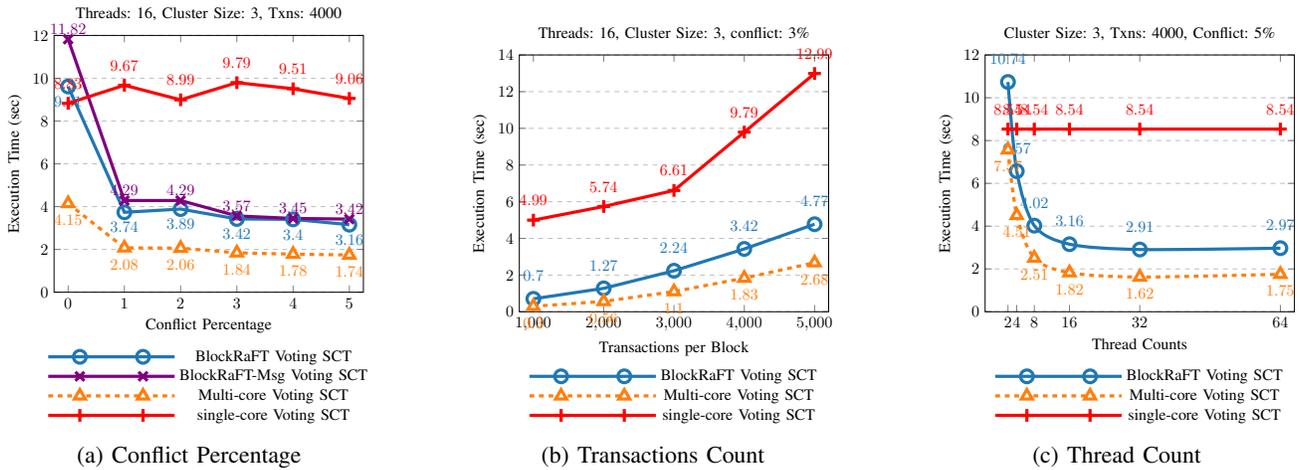

\noindent
\textit{Impact of Thread Count:}
Execution time was measured using 2, 4, 8, 16, 32, and 64 threads with blocks of 4000 transactions. As shown in Fig.~\ref{fig:threads}, increasing the thread count initially improves performance. However, the rate of improvement decreases as the number of threads increases, indicating diminishing returns due to synchronization and coordination overhead. At 64 threads, execution time increases for both BlockRaFT and the multi-core baseline, suggesting contention and parallelization limits.

\begin{figure}[ht]
    \centering
    \begin{tikzpicture}[scale=0.6, every node/.append style={scale=1}]
    \begin{axis}[
        title={Breakdown of Execution Time (Threads: 16, Cluster Size: 3, conflict: 3\%)},
        xlabel={Transactions per Block},
        ylabel={Time (ms)},
        scaled ticks=false,
        xmin=0.5,
        xmax=5.5,
        xtick=data,
        xticklabels={1000,2000,3000,4000,5000},
        ymin=0,
        ymajorgrids=true,
        grid style=dashed,
        enlarge x limits=0.05,
legend style={ at={(0.5,-0.2)}, anchor=north, legend columns=3, fill=none, draw=none, },
        every axis plot/.append style={line width=1.2pt},
        ybar stacked,
        bar width=15pt,
    ]

    \addplot[fill=blue!30, draw=none,]  coordinates {(1,27.0) (2,49.8) (3,40.0) (4,35.0) (5,40.3)};
    \addplot[fill=orange!50, draw=none]  coordinates {(1,36.8) (2,54.2) (3,60.8) (4,90.8) (5,86.7)};
    \addplot[fill=green!40, draw=none]  coordinates {(1,19.8) (2,56.0) (3,125.0) (4,189.6) (5,302.1)};
    \addplot[fill=purple!40, draw=none]  coordinates {(1,7.8) (2,8.6) (3,9.2) (4,10.0) (5,10.1)};
    \addplot[fill=red!50, draw=none]  coordinates {(1,288.6) (2,304.6) (3,287.2) (4,344.6) (5,295.4)};
    \addplot[fill=gray!50, draw=none]  coordinates {(1,262.6) (2,568.6) (3,1119.2) (4,1713.4) (5,2436.6)};

    \legend{
        ETCD Overhead,
        Block Production,
        Component Detection,
        Assigning Followers,
        State Changes,
        Execution Time
    }

    \end{axis}
    \end{tikzpicture}
    \caption{Breakdown of Execution Time}
    \label{fig:breakdown1}
    
\vspace{-2em}
\end{figure}
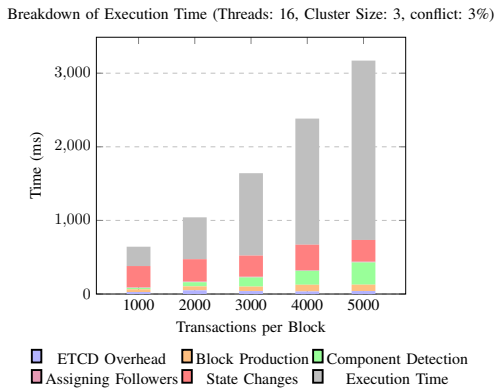

\noindent
\textit{Impact of Conflict Percentage:}
We evaluated performance under transaction dependency levels ranging from 0\% to 5\% using blocks containing 4000 transactions (Fig.~\ref{fig:conflict}). At 0\% dependency, performance is not optimal despite the absence of logical conflicts. This behavior results from the large number of unique addresses, which increases inter-node data transfer and exposes communication as a bottleneck. This trend is further illustrated in the breakdown analysis shown in Fig.~\ref{fig:breakdown1}. 

The BlockRaFT-Msg configuration replaces ETCD with direct message passing for sharing state updates. Although this reduces coordination overhead, communication remains a limiting factor. In future work, we plan to reduce this communication bottleneck to further enhance scalability.

\noindent
\textit{Impact of Transactions per Block:}
We varied the number of transactions per block from 1000 to 5000 under low-conflict conditions. As shown in Fig.~\ref{fig:txns}, execution time increases approximately linearly with transaction volume across all configurations. The single-core baseline exhibits the highest growth rate and longest execution times. In contrast, multi-core and multi-node configurations scale more efficiently, particularly under heavier workloads. Although the multi-node setup introduces coordination overhead at lower workloads, this overhead remains modest while providing the additional benefit of crash fault tolerance.

Based on our experimental evaluation, both research questions are positively addressed. For RQ2 (Scalability), the results demonstrate that BlockRaFT maintains near-linear growth in execution time as transaction volume increases and achieves substantial performance improvements over the baseline models. While diminishing returns appear at higher thread counts due to synchronization overhead, the framework scales efficiently under varying workload and conflict conditions.

For RQ3 (Fault Tolerance), the additional overhead introduced by distributed coordination remains moderate, particularly at higher workloads where multi-node execution outperforms the single-core setup. This confirms that the performance cost of coordination is a reasonable trade-off for the crash fault-tolerance guarantees provided by BlockRaFT.

\paragraph{Impact of Node Failures}
To evaluate fault tolerance, we introduced 1, 2, and 3 node failures in the distributed cluster, maintaining 16 threads and 5\% conflict percentage in the voting smart contract.

\textbf{RQ 4:}  \emph{Does BlockRaFT maintain acceptable performance under node failures?}

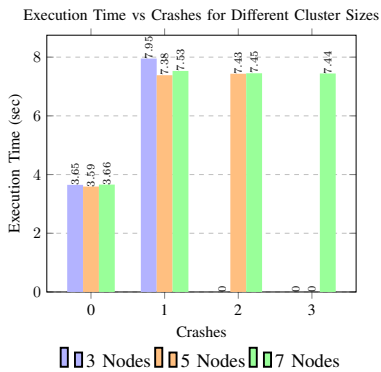
\begin{figure}[ht]
    \centering
    \begin{tikzpicture}[scale=0.6, every node/.append style={scale=1}]
    \begin{axis}[
        title={Execution Time vs Crashes for Different Cluster Sizes},
        xlabel={Crashes},
        ylabel={Execution Time (sec)},
        scaled ticks=false,
        xmin=0,
        xmax=3,
        xtick=data,
        xticklabels={0,1,2,3},
        ymin=0,
        ymajorgrids=true,
        grid style=dashed,
        enlarge x limits=0.2,
        legend style={
            at={(0.5,-0.2)},
            anchor=north,
            legend columns=3,
            fill=none, draw=none,
            legend image post style={scale=1.5},
            font=\large,
        },
        every axis plot/.append style={line width=1.2pt},
        ybar,
        bar width=10pt,
        ybar=0pt,
        nodes near coords,
        every node near coord/.append style={font=\scriptsize, rotate=90, anchor=west, xshift=-0.9ex}
    ]

    \addplot[fill=blue!30, draw=none] coordinates {(0,3.6488) (1,7.952) (2,0) (3,0)};

    \addplot[fill=orange!50, draw=none] coordinates {(0,3.5906) (1,7.3838) (2,7.4345) (3,0)};

    \addplot[fill=green!40, draw=none] coordinates {(0,3.6580) (1,7.5300) (2,7.4510) (3,7.4436)};
    

    \legend{3 Nodes, 5 Nodes, 7 Nodes}

    \end{axis}
    \end{tikzpicture}
    \caption{Execution Time Varying cluster size with crashes}
    \label{fig:crashes}
    \vspace{-1em}
\end{figure}

 As shown in \figref{crashes}, execution time increases after the first failure due to workload increase per system. However, the system continues to operate correctly as long as quorum is maintained. Notably, the performance impact of additional failures beyond the first is comparatively small, indicating that the system stabilizes after the initial redistribution overhead.

These results demonstrate that BlockRaFT maintains functional correctness and exhibits graceful degradation under crash scenarios.

\paragraph{Bottleneck Analysis}

\textbf{RQ 5:}  \emph{What are the bottlenecks in the current distributed framework?}

As shown in Fig. \ref{fig:breakdown1}, the total execution time increases sharply with the number of transactions per block. The main bottlenecks of the current distributed framework are the execution phase and the component detection mechanism, especially under high transaction loads. In future work, we plan to further optimize the execution layer and improve the efficiency of the component detection process to enhance scalability under high transaction loads. The values for breakdown plots and extended experiments are available in the repo \cite{BlockRAFT}

\section{Related Work}
\label{sec:related-work}


Performance limitations remain one of the key obstacles preventing blockchain technologies from being adopted at scale~\cite{Dickerson+:ACSC:PODC:2017}. In response, a variety of research for enhancement techniques have been well explored mainly: parallelized transaction processing, sharding mechanisms, DAG-based Blockchains and distributed transaction execution.

Parallel execution of smartcontracts in Blockchains comes with its own challenges apart from traditions parallel trasaction execution in traditional databases. This avenue for improving throughtput has been well explored in literatures and extensively studied~\cite{Manaswini+:2023:europar,parwat:pdp:2019,yaron+:2024:BSE:SRDS}. ParBlockchain \cite{ParBlock:ICDCS:2019} precomputes a transaction dependency graph and executes only independent transactions in parallel deterministically. Block-STM \cite{blockSTM:2022} similarly accelerates smart-contract execution through optimistic concurrency control with versioned state validation.

Several blockchain architectures have adopted directed acyclic graph (DAG) structures to address these limitations and replace the conventional chain-based model \cite{Qin+:2023:Sok:ACM}. OTA's Tanglev\cite{Hellani+:2021:WETICE:Tangle} treats each transaction as a node in a DAG, requiring new transactions to approve two previous transactions. PHANTOM \cite{Sompolinsky+:2021:Phantom:AFT} extends the proof of work \cite{Nakamoto:Bitcoin:2009} consensus to a blockDAG,  distinguishes between blocks mined properly by honest nodes and those created by non-cooperating nodes who choose to deviate from the mining protocol. Adopting DAG-based blockchain systems also implies that scalability is no longer constrained by block size, shifting the focus toward optimizing node performance and network-level concurrency to handle increased transaction loads effectively. Beyond sharding \cite{Dang+:SIGMOD:2019, loi+:shard:CCS:2016} and parallelism, researchers have explored distributed blockchain architectures that improve throughput through workload distribution. 
\begin{table}[h!]
\centering
\begin{tabular}{|l|c|c|c|c|}
\hline
\textbf{Feature / System} & \textbf{DiPETrans} & \textbf{PilotFish} & \textbf{BlockRaft} & \textbf{Sharding} \\ \hline

\textbf{Scalable} & \cmark & \cmark & \cmark & \cmark \\ \hline

\textbf{Node Crash} & \xmark & \xmark & \cmark & \xmark \\ 
\textbf{Tolerance} &  &  &  &  \\\hline

\textbf{Workload} & \xmark & \xmark & \cmark & \xmark \\ 
\textbf{Distribution} &  &  &  &  \\\hline

\textbf{Distributed} & \xmark & \cmark & \cmark & \cmark \\

\textbf{SCTs Execution}  &  &  &  &  \\\hline

\textbf{Parallel Mining} & \cmark & \xmark & \xmark & \xmark \\ \hline

\textbf{Trustless} & \xmark & \cmark & \cmark & \cmark \\
\textbf{Operation} &  &  &  &  \\\hline

\textbf{Compatibility} & \cmark & \cmark & \cmark & \xmark \\ \hline

\textbf{Blockchain} & \xmark & \cmark & \cmark & \cmark \\
\textbf{Security Impact} &  &  &  &  \\\hline

\end{tabular}
\caption{Comparison of DiPETrans~\cite{DiPETrans:CPE:2022}, PilotFish~\cite{Kniep+:2025:pilotfish:FC}, BlockRaft, and general sharding-based~\cite{loi+:shard:CCS:2016,Pengze+:2024:ACM:spring} architectures}
\label{tab:comparison}
\end{table}

DiPETrans \cite{DiPETrans:CPE:2022} presents a framework for parallelizing transaction execution within a block by leveraging a community of peers. Using static analysis, a leader node groups independent transactions into shards and assigns them to follower nodes for parallel execution. It employs a transaction ordering service to establish a global sequence of transactions and delegates execution to distributed worker nodes. Both mining and validation are parallelized, leveraging community compute power.

PilotFish \cite{Kniep+:2025:pilotfish:FC} introduces a three-layer system architecture for a distributed execution engine consisting of a global transaction sequencer, a set of execution workers, and distributed storage. Transactions are processed across execution nodes in a pipelined fashion, and execution consensus maintains consistency without requiring all nodes to re-execute every transaction. This architecture reduces redundant computation and supports high throughput in permissioned blockchain settings. Pilotfish demonstrated linear scaling of an eightfold throughput increase with eight machines, while maintaining low latency and avoiding the batching delays.

In this work, we concentrate on workload division for not just execution but also all blockchain node operations. Improving the execution is insufficient to improve blockchain scalability because one should look at all the operations blockchain nodes perform. We also explore crash tolerance to decrease the load on the blockchain network caused by the replication solution for crash tolerance. The comparison in Table~\ref{tab:comparison} highlights key differences among DiPETrans, PilotFish, BlockRaft, and general sharding-based architectures. While all systems demonstrate scalability, their levels of resilience and decentralization vary notably. BlockRaft is the only architecture that provides strong crash tolerance, whereas DiPETrans, PilotFish, and common sharding designs do not inherently tolerate node failures. BlockRaft also uniquely distributes the full blockchain workload, unlike the others. Parallel mining is exclusive to DiPETrans, as the remaining systems are not oriented around mining tasks. Trust models further differentiate the systems: PilotFish, BlockRaft, and sharding operate trustlessly, while DiPETrans relies on a trusted community. BlockRaft also integrating seamlessly into existing systems, whereas sharding typically requires architectural changes.

\section{Conclusion}

In this paper, we presented BlockRaFT, a crash-tolerant and scalable distributed framework for blockchain nodes. Our approach uses a leader-follower model to distribute workloads efficiently and applies storage optimizations to improve smart contract performance. Additionally, we incorporate data storage optimizations that are particularly effective for handling smart contract operations, further enhancing system performance. We implemented and evaluated our framework through a series of experiments, demonstrating that BlockRaFT consistently outperforms single-core implementations in terms of throughput and responsiveness. BlockRaFT introduces only a small overhead compared to multicore setups, which is a reasonable trade-off considering its improved crash tolerance and distributed resilience. We further propose a Merkle tree optimization that decouples database updates from smart contract execution.

As part of our future work, we aim to optimize BlockRaFT’s performance by introducing a data sharing mechanism. This is motivated by our observation of overhead introduced by data sharing, particularly when operating at 0\% conflict percentage. We also intend to explore a more decentralized, peer-to-peer architecture. 


\bibliographystyle{IEEEtran}
\bibliography{citations}
\newpage
\appendix
\label{apn:appendix}

\subsection{Implementation Details}

We have implemented the proposed BlockRaFT framework using the C++ programming language. We utilize ETCD \cite{etcd}, which serves as a distributed, asynchronous shared memory space for asynchronous communication and coordination among cluster nodes. ETCD simplifies coordination tasks by allowing nodes to exchange state information consistently. In \figref{dist} we illustrate the storage structure that we use for cluster node communication.

\begin{figure}[h]
  \centering
        \centering
        \includegraphics[scale=0.24]{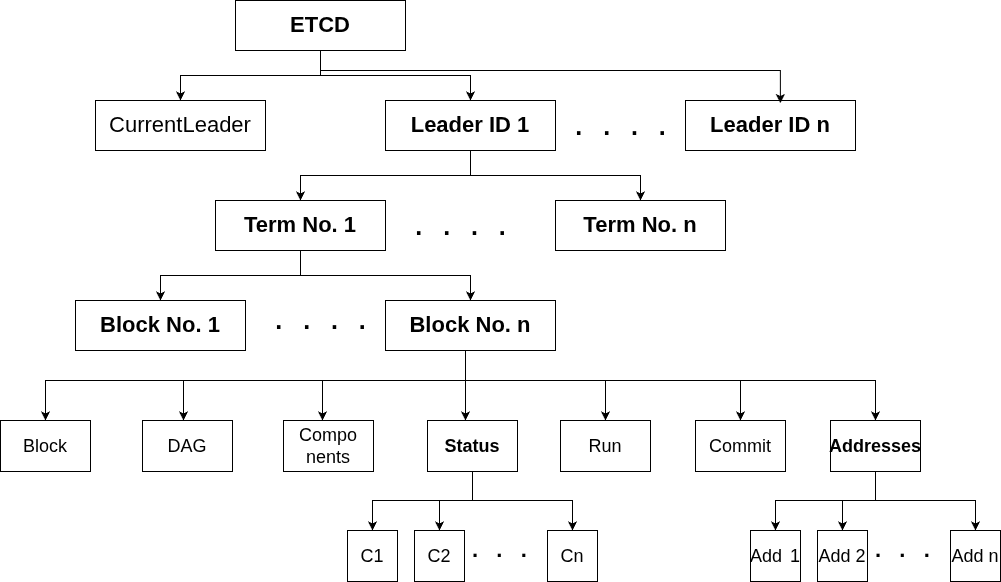}
        \caption{ETCD storage structure.}
	\label{fig:etcd}
\end{figure}

Since ETCD employs the RAFT consensus algorithm internally, our framework leverages its built-in leader election mechanism. Consequently, the node designated as the leader by ETCD is also considered the leader in our transaction execution framework. To track the leader’s availability, we make use of ETCD’s lease and time-to-live (TTL) mechanisms to signal the activeness of the leader node. Additionally, follower nodes employ ETCD’s watcher function to monitor leadership changes in real time.

System health is proactively monitored across the cluster: all nodes continuously check the health of both the ETCD cluster and the Redpanda transaction queue. For maintaining a globally shared transaction pool, we employ a distributed, asynchronous multi-producer, single-consumer queue architecture. In this setup, all nodes' network layer and REST APIs act as producers by continuously submitting transactions, while the leader node serves as the sole consumer during block production. For this purpose, we incorporate Redpanda \cite{redpanda}, a high-performance streaming platform, to efficiently manage and maintain the transaction pool. Furthermore, the number of threads allocated to both the DAG module and the parallel scheduler are configurable and specified via a system configuration file, allowing fine-grained control over performance tuning based on deployment requirements.

\subsection{Results}
\label{sec:wallet}
This appendix presents a detailed breakdown of five experiments evaluating the performance of the BlockRaFT system under varying conditions. The key metrics observed include conflict percentage, number of threads, transaction volume, and fault tolerance across different SCT (Smart Contract Type) workloads.

\subsubsection{Wallet Smart Contract}
This contract represents a decentralized banking or token-based system, where wallet balances and fund transfers are securely recorded and verifiable through the blockchain's global state. The transactions under this SCT are:
\begin{itemize}
  \item \texttt{deposit <client\_id> <key> <amount>}: Adds funds to a wallet.
  \item \texttt{withdraw <client\_id> <key> <amount>}: Removes funds from a wallet.
  \item \texttt{transfer <from\_id> <from\_key> <to\_id> <to\_key> <amount>}: Transfers funds between wallets.
  \item \texttt{balance <client\_id> <key>}: Displays the current wallet balance.
\end{itemize}

\begin{table*}[h!]
\centering
\caption{Exp 8. Conflict Percentage Breakdown for Voting SCTs (Threads:16, Txn:4000, Cluster:3)}
\begin{tabular}{|c|c|c|c|c|c|c|}
\hline
\textbf{Conflict} & \textbf{ETCD} & \textbf{Block} & \textbf{Component } & \textbf{Assigning } & \textbf{Storing } & \textbf{Execution } \\
\textbf{Percentage} & \textbf{Ops (ms)} & \textbf{ Production (ms)} & \textbf{ Detection (ms)} & \textbf{ Followers (ms)} & \textbf{Value (ms)} & \textbf{ Time (ms)} \\\hline
0 & 40.8 & 76.0 & 155.8 & 10.0 & 6595.6 & 1796.0 \\ \hline
1 & 37.8 & 70.6 & 172.4 & 9.4  & 666.8  & 1709.6 \\ \hline
2 & 34.4 & 78.6 & 193.0 & 8.8  & 791.2  & 1752.6 \\ \hline
3 & 35.0 & 90.8 & 189.6 & 10.0 & 344.6  & 1713.4 \\ \hline
4 & 38.4 & 77.6 & 209.0 & 9.0  & 338.4  & 1738.8 \\ \hline
5 & 49.0 & 72.8 & 221.2 & 11.0 & 181.8  & 1626.6 \\ \hline
\end{tabular}
\end{table*}

\begin{table*}[h!]
\centering
\caption{Exp 9. Transactions per Block breakdown for Voting SCTs (Threads:16, Cluster:3, Clonflict:3\%)}
\begin{tabular}{|c|c|c|c|c|c|c|}
\hline
\textbf{Txns per} & \textbf{ETCD} & \textbf{Block} & \textbf{Component } & \textbf{Assigning } & \textbf{Storing } & \textbf{Execution } \\
\textbf{Block} & \textbf{Ops (ms)} & \textbf{ Production (ms)} & \textbf{ Detection (ms)} & \textbf{ Followers (ms)} & \textbf{Value (ms)} & \textbf{ Time (ms)} \\\hline
1000 & 27.0 & 36.8 & 19.8  & 7.8  & 288.6 & 262.6  \\ \hline
2000 & 49.8 & 54.2 & 56.0  & 8.6  & 304.6 & 568.6  \\ \hline
3000 & 40.0 & 60.8 & 125.0 & 9.2  & 287.2 & 1119.2 \\ \hline
4000 & 35.0 & 90.8 & 189.6 & 10.0 & 344.6 & 1713.4 \\ \hline
5000 & 40.3 & 86.7 & 302.1 & 10.1 & 295.4 & 2436.6 \\ \hline
\end{tabular}
\end{table*}

\begin{table*}[h!]
\centering
\caption{Exp 1. Thread count Vs Execution time (ms) for Wallet SCTs (Txn:4000, Cluster:3, Clonflict:5\%)}
\begin{tabular}{|c|c|c|c|}
\hline
\textbf{No of Threads} & \textbf{Multi-node (ms)} & \textbf{Multi-core (ms)} & \textbf{Single-core (ms)} \\
\hline
2  & 10744  & 7671.2 & 8452.2 \\ \hline
4  & 6699.4 & 4628.4 & 8452.2 \\ \hline
8  & 4061.4 & 2586.2 & 8452.2 \\ \hline
16 & 3272.8 & 1883.6 & 8452.2 \\ \hline
32 & 2985.8 & 1703   & 8452.2 \\ \hline
64 & 2944   & 1779.6 & 8452.2 \\ \hline
\end{tabular}
\end{table*}

\begin{figure}[ht]
    \centering
    \begin{tikzpicture}[scale = 0.9,every node/.append style = {scale = 1}]
    \begin{axis}[
    title={Execution Time vs Thread Counts (Cluster Size: 3, Txns: 4000, Conflict: 5\%)},
    xlabel={Thread Counts},
    ylabel={Execution Time (sec)},
    scaled ticks=false,
    xmin=0, xmax=70,
    xtick={2,4,8,16,32,64},
    ymin=0, ymax=12,
    ytick={0,2,4,6,8,10,12},  
    ymajorgrids = true,
    grid style = dashed,
    enlarge x limits = 0.05,
    legend style = {
        at = {(0.5,-0.25)},
        anchor = north,
        legend columns = 1,
        fill=none, draw=none,
    },
    every axis plot/.append style = { line width = 2pt },
    nodes near coords,
    ]

    \addplot[color=walletcolor, mark=o, mark options={fill=white,solid}, mark size=4pt, smooth]
        coordinates {(2,10.744) (4,6.6994) (8,4.0614) (16,3.2728) (32,2.9858) (64,2.9440)};
        
    \addplot[color=mixedcolor, mark=triangle*, mark options={fill=white,solid}, mark size=4pt, dashed, smooth]
        coordinates {(2,7.6712) (4,4.6284) (8,2.5862) (16,1.8836) (32,1.7030) (64,1.7796)};
        
    \addplot[color=red, mark=+, mark options={fill=white,solid}, mark size=4pt, solid, smooth]
        coordinates {(2,8.4522) (4,8.4522) (8,8.4522) (16,8.4522) (32,8.4522) (64,8.4522)};
    
    \legend{Multi-node Wallet SCT, Multi-core Wallet SCT, Single-core Wallet SCT}

    \end{axis}	
\end{tikzpicture}
\caption{Performance Comparison across Thread Counts}
\end{figure}
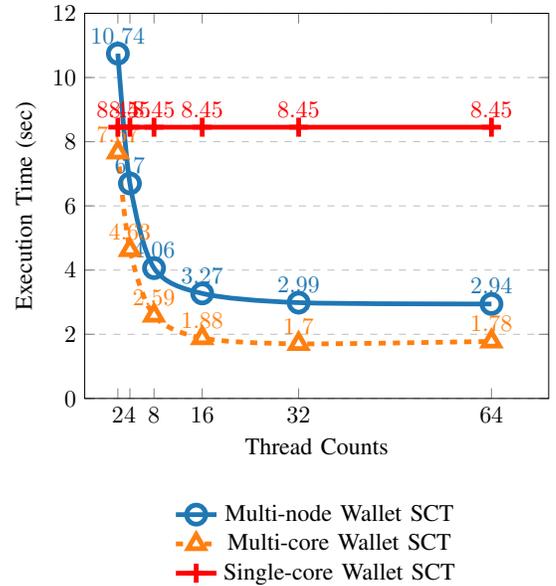

\begin{table*}[h!]
\centering
\caption{Exp 2. Conflict Percentage (0--100) Vs Execution time (ms) for Wallet SCTs (Threads:16, Txn:4000, Cluster:3)}
\begin{tabular}{|c|c|c|c|}
\hline
\textbf{Conflict Percentage} & \textbf{Multi-node (ms)} & \textbf{Multi-core (ms)} & \textbf{Single-core (ms)} \\
\hline
0   & 10453.6 & 3504.2 & 9965.6 \\ \hline
20  & 3590.2  & 1707   & 9415.6 \\ \hline
40  & 3510.2  & 1555.2 & 9311.6 \\ \hline
60  & 3462.4  & 1359.6 & 9031.4 \\ \hline
80  & 3280.6  & 1150.4 & 7370.8 \\ \hline
100 & 3204.0  & 906.8  & 7365.6 \\ \hline
\end{tabular}
\end{table*}


\begin{figure}[ht]
    \centering
    \begin{tikzpicture}[scale=0.9]
    \begin{axis}[
    title={Execution Time vs Conflict Percentage},
    xlabel={Conflict Percentage (\%)},
    ylabel={Execution Time (sec)},
    xmin=0, xmax=100,
    xtick={0,20,40,60,80,100},
    ymin=0, ymax=12,
    ytick={0,2,4,6,8,10,12},
    ymajorgrids=true,
    grid style=dashed,
    legend style={at={(0.5,-0.25)},anchor=north,legend columns=1,fill=none,draw=none},
    every axis plot/.append style={line width=2pt},
    nodes near coords,
    ]
    \addplot[color=walletcolor,mark=o,mark options={fill=white,solid},mark size=4pt]
        coordinates {(0,10.4536) (20,3.5902) (40,3.5102) (60,3.4624) (80,3.2806) (100,3.2040)};
    \addplot[color=mixedcolor,mark=triangle*,mark options={fill=white,solid},mark size=4pt,dashed]
        coordinates {(0,3.5042) (20,1.7070) (40,1.5552) (60,1.3596) (80,1.1504) (100,0.9068)};
    \addplot[color=red,mark=+,mark options={fill=white,solid},mark size=4pt,solid]
        coordinates {(0,9.9656) (20,9.4156) (40,9.3116) (60,9.0314) (80,7.3708) (100,7.3656)};
    \legend{Multi-node Wallet SCT, Multi-core Wallet SCT, Single-core Wallet SCT}
    \end{axis}
    \end{tikzpicture}
    \caption{Performance Comparison across Conflict Percentages}
\end{figure}

\begin{table*}[h!]
\centering
\caption{Exp 3. Conflict Percentage (0--5) Vs Execution time (ms) for Wallet SCTs (Threads:16, Txn:4000, Cluster:3)}
\begin{tabular}{|c|c|c|c|}
\hline
\textbf{Conflict Percentage} & \textbf{Multi-node (ms)} & \textbf{Multi-core (ms)} & \textbf{Single-core (ms)} \\
\hline
0 & 10453.6 & 3504.2 & 9965.6 \\ \hline
1 & 3934.4  & 1966.4 & 9179.2 \\ \hline
2 & 4120.6  & 2079.2 & 9683.4 \\ \hline
3 & 3888.6  & 1909.4 & 8966.8 \\ \hline
4 & 3508.6  & 1877.8 & 9533.4 \\ \hline
5 & 3414.4  & 1802.6 & 9846.6 \\ \hline
\end{tabular}
\end{table*}

\begin{figure}[ht]
    \centering
    \begin{tikzpicture}[scale=0.9]
    \begin{axis}[
    title={Execution Time vs Conflict Percentage},
    xlabel={Conflict Percentage (\%)},
    ylabel={Execution Time (sec)},
    xmin=0, xmax=5,
    xtick={0,1,2,3,4,5},
    ymin=0, ymax=12,
    ytick={0,2,4,6,8,10,12},
    ymajorgrids=true,
    grid style=dashed,
    legend style={at={(0.5,-0.25)},anchor=north,legend columns=1,fill=none,draw=none},
    every axis plot/.append style={line width=2pt},
    nodes near coords,
    ]
    \addplot[color=walletcolor,mark=o,mark options={fill=white,solid},mark size=4pt]
        coordinates {(0,10.4536) (1,3.9344) (2,4.1206) (3,3.8886) (4,3.5086) (5,3.4144)};
    \addplot[color=mixedcolor,mark=triangle*,mark options={fill=white,solid},mark size=4pt,dashed]
        coordinates {(0,3.5042) (1,1.9664) (2,2.0792) (3,1.9094) (4,1.8778) (5,1.8026)};
    \addplot[color=red,mark=+,mark options={fill=white,solid},mark size=4pt,solid]
        coordinates {(0,9.9656) (1,9.1792) (2,9.6834) (3,8.9668) (4,9.5334) (5,9.8466)};
    \legend{Multi-node Wallet SCT, Multi-core Wallet SCT, Single-core Wallet SCT}
    \end{axis}
    \end{tikzpicture}
    \caption{Performance Comparison across Conflict Percentages}
\end{figure}

\begin{table*}[h!]
\centering
\caption{Exp 4. Txn Count Vs Execution time (ms) for Wallet SCTs (Threads:16, Cluster:3, Clonflict:3\%)}
\begin{tabular}{|c|c|c|c|}
\hline
\textbf{Txns / block} & \textbf{Multi-node (ms)} & \textbf{Multi-core (ms)} & \textbf{Single-core (ms)} \\
\hline
1000 & 770    & 279.8  & 4103   \\ \hline
2000 & 1316.4 & 614.4  & 5727.8 \\ \hline
3000 & 2231.4 & 1132.2 & 7618.2 \\ \hline
4000 & 3888.6 & 1844.8 & 8966.8 \\ \hline
5000 & 5083   & 2772.6 & 12236.8 \\ \hline
\end{tabular}
\end{table*}

\begin{figure}[ht]
    \centering
    \begin{tikzpicture}[scale = 0.8,every node/.append style = {scale = 1}]
    \begin{axis}[
    title={Execution Time vs Transactions per Block (Threads: 16, Cluster Size: 3, Conflict: 3\%)},
    xlabel={Transactions per Block},
    ylabel={Execution Time (sec)},
      scaled ticks=false,
         xmin=1000,
         xmax=5000,
    xtick=data,
     ymin=0,
     ymax=14,
    ytick={0,2,4,6,8,10,12,14},
        ymajorgrids = true,
        grid style = dashed,
        enlarge x limits = 0.05,
        legend style = {
            at = {(0.5,-0.25)},
            anchor = north,
            legend columns = 1,
            fill=none, draw=none,
        },
        every axis plot/.append style = { line width = 2pt },
        /pgfplots/legend image code/.code = {%
            \draw[mark repeat=2,mark phase=2,#1] 
            plot coordinates {
                (0cm,0cm) 
                (0.7cm,0cm)
                (1.4cm,0cm)
                (2.1cm,0cm)
                (2.8cm,0cm)%
            };
        },
        nodes near coords,
    ]

        \addplot[color=walletcolor, mark=o, mark options={fill=white,solid}, mark size=4pt,
                 every node near coord/.append style = {xshift = 0pt, yshift = 7pt}] 
            table[x=txns, y=BlockRAFT, col sep=space] {results/txns.txt};

        \addplot[color=mixedcolor, mark=triangle*, mark options={fill=white,solid}, mark size=4pt, dashed,
                 every node near coord/.append style = {xshift = 0pt, yshift = -18pt}] 
            table[x=txns, y=MultiCore, col sep=space] {results/txns.txt};

        \addplot[color=red, mark=+, mark options={fill=white,solid}, mark size=4pt, solid,
                 every node near coord/.append style = {xshift = 0pt, yshift = 5pt}] 
            table[x=txns, y=SingleCore, col sep=space] {results/txns.txt};

        \legend{BlockRaFT Wallet SCT, Multi-core Wallet SCT, Single-core Wallet SCT}

    \end{axis}	
\end{tikzpicture}

    \caption{Performance Comparison across Transactions per Block}
    \label{fig:txns}
\end{figure}
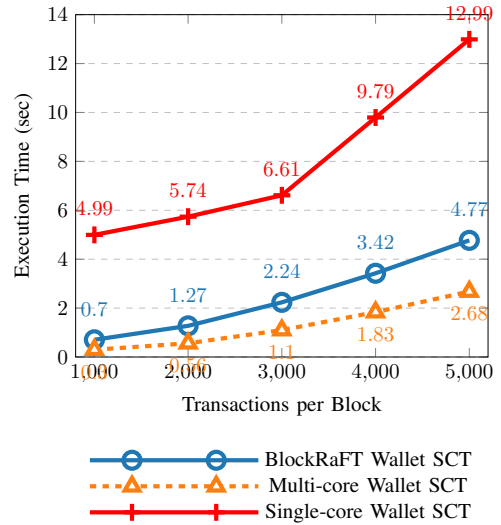

\begin{table*}[h!]
\centering
\caption{Exp 6. Txn Count Vs Execution time (ms) for Wallet SCTs (Threads:16, Txn:4000, Conflict:1\%)}
\begin{tabular}{|c|c|}
\hline
\textbf{Cluster Size} & \textbf{Execution time (ms)} \\
\hline
3 & 3934.4 \\ \hline
5 & 3621.6 \\ \hline
7 & 3817.8 \\ \hline
\end{tabular}
\end{table*}

\begin{table*}[h!]
\centering
\caption{Exp 7. Crashes Vs Execution time (ms) for Wallet SCTs (Threads:16, Txn:4000, Conflict:1\%)}
\begin{tabular}{|c|c|c|c|}
\hline
\textbf{Crashes} & \textbf{3 Nodes} & \textbf{5 Nodes} & \textbf{7 Nodes} \\
\hline
0 & 3934.4  & 3621.6  & 3817.8  \\ \hline
1 & 7890.8  & 7524.25 & 7539.2  \\ \hline
2 & N/A     & 7412.5  & 7546    \\ \hline
3 & N/A     & N/A     & 7423    \\ \hline
\end{tabular}
\end{table*}

\textbf{Experiment 1: Thread Count vs Execution Time (ms)}
Parameters and Values for Experiment 1
\begin{itemize}
    \item \textbf{Conflict}: The conflict percentage is 5\%.
    \item \textbf{Cluster Size}: The cluster size is 3.
    \item \textbf{Txn}: The number of transactions is 4000.
\end{itemize}
This experiment evaluates how execution time scales with the number of threads. The system ran 4000 transactions at 5\% conflict with a cluster size of 3. Performance initially degrades with increased threads due to concurrency overhead. However, with 16 threads, throughput significantly improves, particularly in the wallet and voting SCTs. Notably, the wallet multi-core variant shows the highest scalability, suggesting better utilization of thread parallelism.

\textbf{Experiment 2: Conflict percentatio (0\% to 100\%) vs Execution Time (ms)}

Parameters and Values for Experiment 2
\begin{itemize}
    \item \textbf{Threads}: The number of threads used is 16.
    \item \textbf{Cluster Size}: The cluster size is 3.
    \item \textbf{Txn}: The number of transactions is 4000.
\end{itemize}
Here, we fix the thread count at 16 and vary the conflict percentage from 0\% to 100\%, simulating contention-heavy workloads.  As expected, performance declines as conflict increases. At 0\% conflict, the BlockRaft performs poorly due to communication overhead.

\textbf{Experiment 4: No. of Transactions vs Execution Time (ms)}

Parameters and Values for Experiment 4
\begin{itemize}
    \item \textbf{Threads}: The number of threads used is 16.
    \item \textbf{Conflict}: The conflict percentage is 2\%.
    \item \textbf{Cluster Size}: The cluster size is 3.
    We evaluate performance under increasing transaction counts, from 1000 to 5000, with a fixed 2\% conflict and 16 threads. Throughput scales approximately linearly, although wallet single-core shows limited scalability. BlockRaFT variants, particularly voting SCT, handle larger transaction loads effectively.
\end{itemize}

\textbf{Experiment 5: No. of Crashes vs Execution Time (ms)}

Parameters and Values for Experiment 5
\begin{itemize}
    \item \textbf{Threads}: The number of threads used is 16.
    \item \textbf{Conflict}: The conflict percentage is 5\%.
    \item \textbf{SCT}: Voting SCT
    \end{itemize}

This experiment tests resilience by inducing node crashes in clusters of 3, 5, and 7 nodes under 5\% conflict and 16 threads, focusing on the voting SCT.

\end{document}